\documentclass[9.5pt,journal,compsoc]{IEEEtran}

\usepackage{tcolorbox}
\usepackage{caption}
\usepackage[colorinlistoftodos]{todonotes}
\usepackage{lipsum}
\usepackage{booktabs}
\usepackage{multirow}
\usepackage{supertabular}
\usepackage{supertabular,booktabs}
\usepackage{amsmath}
\usepackage[inline]{enumitem}
\usepackage{ragged2e}
\usepackage[switch]{lineno}
\usepackage{xcolor}
\usepackage{tabularx}
\usepackage{breakurl}

\ifCLASSOPTIONcompsoc
  \usepackage[nocompress]{cite}
\else
  \usepackage{cite}
\fi

\ifCLASSINFOpdf
\else
\fi

\begin{document}
\title{Mining Temporal Attack Patterns from Cyberthreat Intelligence Reports}
\author{Md Rayhanur Rahman, Brandon Wroblewski, Quinn Matthews, Brantley Morgan, Tim Menzies, Laurie Williams~\thanks{The authors are with the Department of Computer Science, North Carolina State University, Raleigh, NC 27606 USA. E-mail: \{mrahman, bnwroble, qdmatthe, bmmorgan, tjmenzie, lawilli3\}@ncsu.edu.}}

\markboth{Transaction of Software Engineering}%
{Shell \MakeLowercase{\textit{et al.}}: Bare Demo of IEEEtran.cls for Computer Society Journals}

\IEEEtitleabstractindextext{%
\begin{abstract} 

\justifying{Defending from cyberattacks requires practitioners to operate on high-level adversary behavior. Cyberthreat intelligence (CTI) reports on past cyberattack incidents describe the chain of malicious actions with respect to time. To avoid repeating cyberattack incidents, practitioners must proactively identify and defend against recurring chain of actions - which we refer to as temporal attack patterns. Automatically mining the patterns among actions provides structured and actionable information on the adversary behavior of past cyberattacks. \textit{The goal of this paper is to aid security practitioners in prioritizing and proactive defense against cyberattacks by mining temporal attack patterns from cyberthreat intelligence reports}. To this end, we propose \textbf{ChronoCTI}, an automated pipeline for mining temporal attack patterns from cyberthreat intelligence (CTI) reports of past cyberattacks. To construct \textbf{ChronoCTI}, we build the ground truth dataset of temporal attack patterns and apply state-of-the-art large language models, natural language processing, and machine learning techniques. We apply \textbf{ChronoCTI} on a set of 713 CTI reports, where we identify 124 temporal attack patterns - which we categorize into nine pattern categories. We identify that the most prevalent pattern category is to trick victim users into executing malicious code to initiate the attack, followed by bypassing the anti-malware system in the victim network. Based on the observed patterns, we advocate organizations to train users about cybersecurity best practices, introduce immutable operating systems with limited functionalities, and enforce multi-user authentications. Moreover, we advocate practitioners to leverage the automated mining capability of \textbf{ChronoCTI} and design countermeasures against the recurring attack patterns.}

\end{abstract}

}

\maketitle

\IEEEdisplaynontitleabstractindextext

%
\IEEEpeerreviewmaketitle

\newtcolorbox[auto counter]{mybox}[2][]{%
title=Example~\thetcbcounter: #2, #1}

\IEEEraisesectionheading{\section{Introduction}\label{sec:introduction}}

The cost of cyberattacks to organizations is estimated to be \$10.5 trillion annually by 2025~\cite{wef}. Moreover, recent cyberattacks are multi-staged, less opportunistic, and more organized.  The attackers study their victims thoroughly and design attack vectors tailored to their target software systems. Thus, traditional signature-based (i.e., IP addresses, malware hashes) defense mechanisms such as firewalls and malware protection tools do not perform well in detecting and defending these attacks~\cite{ren2022cskg4apt}. However, cybersecurity practitioners can leverage cyberthreat intelligence (CTI)~\cite{rahman2023attackers}, which refers to the information that can aid practitioners in proactively defending themselves from cyberattacks~\cite{mcmillan_definition_nodate}.




Reputed cybersecurity vendors publish technical reports on past cyberattack incidents, commonly known as CTI reports. The reports aid practitioners in understanding high-level adversary behaviors. As attackers vary the technical aspects of cyberattacks, operating on the patterns of attack behaviors instead of low-level attack indicators benefits security practitioners more~\cite{pyramidOfPain}. CTI reports serve as a great source for attack behaviors. Because, the reports contain the temporal relations among attacker actions, which refers to the chain of malicious cyberattack actions with respect to time. Example 1 shows two excerpts from two CTI reports describing two different cyberattacks. In both attacks, \textit{phishing} action is followed by \textit{macro execution} action in terms of time. The example indicates that a recurring sequence of actions exists in multiple cyberattacks. We refer to the recurring temporal sequence as temporal attack patterns. 


However, mining the temporal patterns from unstructured text is challenging because the task requires a deep semantic understanding of the cyberattack incident and the temporal relations among the different malicious actions described in the text. Automatically mining the patterns can benefit practitioners by (a) reducing human effort and time to parse complex CTI reports and (b) providing proactive CTI on how adversaries operate by composing multiple actions. The CTI aids practitioners in thinking like attackers and knowing what malicious actions are likely to happen next - which enables the practitioners to assess imminent threats and strengthen their cybersecurity readiness.

\textit{The goal of this paper is to aid security practitioners in prioritizing and proactive defense against attacks by mining temporal attack patterns from open-source cyberthreat intelligence reports}. We investigate the following research questions (RQs). 

\noindent \textbf{RQ1:} How do we automatically mine temporal attack patterns from cyberthreat intelligence reports? \\
\textbf{RQ2:} What temporal attack patterns do we identify from past incidents of cyberattacks?

\begin{figure*}
    \centering
    \includegraphics[width=\textwidth]{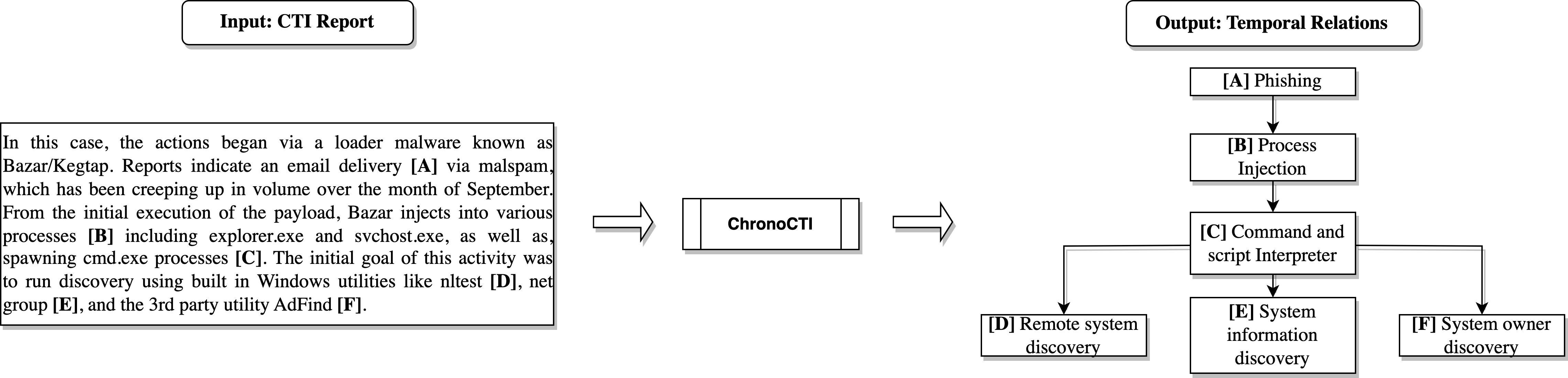}
    \caption{A simplified example of extracted temporal patterns from a CTI report~\cite{dfirReportRyukReturn} by ChronoCTI}
    \label{fig:example}
\end{figure*}

To answer RQ1, we investigate large language models to identify actions and extract temporal features from the text. We construct a ground truth dataset of temporal relations among actions from a set of 94 CTI reports, upon which we train supervised classifiers. We propose a large language model-based machine learning pipeline named \textbf{ChronoCTI} to mine temporal relations from CTI reports. To answer RQ2, we apply \textbf{ChronoCTI} on 713 CTI reports describing past cyberattacks to identify temporal attack patterns. Overall, we contribute the following:
\begin{enumerate*}[leftmargin=4mm]
    \item An automated temporal relation mining pipeline named \textbf{ChronoCTI}. Given a textual description of an attack incident, \textbf{ChronoCTI} outputs the temporal relations among actions. Fig.~\ref{fig:example} shows a simplified example of the capability of ChronoCTI. To the best of our knowledge, \textbf{ChronoCTI} is the first ML pipeline for mining temporal relations among actions from text; 
    \item a ground truth dataset of (a) mapping between sentences and adversary actions, (b) temporal relations among adversary actions of 94 cyberattack incident descriptions, (c) fine-tuned large language models trained on cybersecurity-specific domain corpus. We open-source the dataset and machine learning models so security practitioners and researchers can use, modify, and introduce enhancements to our proposed approach and
    \item a set of 124 temporal patterns across nine pattern categories, derived from 713 CTI reports by \textbf{ChronoCTI}. 
    \item a set of best practices, suggested policies, and countermeasures against the derived patterns
\end{enumerate*}

We organize the rest of the paper as follows. We discuss related concepts in Section~\ref{sec:concept}. We describe the methodology and findings in Section~\ref{sec:method},~\ref{sec:rq1Findings},~\ref{sec:rq2Findings} respectively. We provide further discussions on the findings in Section~\ref{sec:discussion}, followed by related work and conclusion in Section~\ref{sec:relatedWork}, and~ \ref{sec:conclusion}. 

\noindent \fbox{\begin{minipage}{26em}
\textbf{Example 1:} Excerpts from two CTI reports~\cite{dfirReportCollectExfiltrateSleepRepeat}, and \cite{dfirReportIcedIdMalware} showing actions in bold: \textbf{Excerpt 1:} [...] we observed a compromise that was initiated with a Word \textbf{document} containing a malicious VBA \textbf{macro}, which established persistence and communication to a command and control server (C2). \textbf{Excerpt 2: } [...] this document was delivered as part of a \textbf{malicious email campaign} [...]. Upon opening the Excel document, the \textbf{macros would be executed} [...]. 

\end{minipage}}

\section{Key Concepts}
\label{sec:concept}
We discuss several key concepts in this section. 

\textbf{MITRE ATT\&CK:} To model attacks through uniform terminologies, MITRE corporation introduced ATT\&CK~\cite{mitreAttack} in 2013 - which enumerates the cybercrime groups and malware and techniques used by them. For each case of a group or malware using technique(s), ATT\&CK contains a link to the CTI reports where an instance of a technique usage is documented - which we refer to as a \textit{citation}. 


\textbf{Tactics, Techniques, and Procedures (TTPs):} To represent adversary action, ATT\&CK uses a term called \textit{tactics, techniques, and procedures (TTPs)}~\cite{techniqueDefNIST,techniqueDefFeroot}. \textit{Tactic} refers to the goal of an adversary for performing a malicious action. For example, \textit{credential access}~\cite{TA0006} is a tactic where the adversary obtains access to victim credentials. \textit{Techniques} refer to an abstract action through which adversary fulfill the tactic~\cite{mitreAttack, attack-design}. For example, the adversary can fulfill \textit{credential-access} tactic via capturing user input, known as \textit{T1056: Input Capture}~\cite{T1056} in ATT\&CK. \textit{Procedures} refer to the textual description of how the technique is implemented~\cite{attack-design}. For example, "FlawedAmmyy can collect keyboard events"~\cite{flawwedAmmy} - described as a procedure for the \textit{input capture} technique. ATT\&CK contains an enumeration of tactics, where each tactic has an enumeration of techniques and an enumeration of related procedure(s). In ATT\&CK, the identifier of a technique takes the form: \textit{Txxxx}. A technique may have sub-techniques, and the identifier of the sub-technique follows the convention of the identifier of a technique, followed by \textit{.xxx}. For example, the id of the \textit{Input Capture} technique is \textit{T1056}. The definition of all techniques and sub-techniques are available at the official ATT\&CK website.~\footnote{documentation of a technique with an Id \textit{Txxxx} can be found at https://attack.mitre.org/techniques/Txxxx/. E.g: Documentation of T1056: Input Capture can be found at https://attack.mitre.org/techniques/T1056/}  

\textbf{Temporal Relation between Attack Action:} In cyberattacks, adversaries perform malicious actions step by step as time progresses. For instance, in Example 1, we see that after a phishing email is sent, a victim opens the document and executes the malicious macro. Here, phishing action is followed by macro execution action. The example indicates a temporal relation between \textit{phishing} and \textit{macro execution} action. Academic literature already contains a well-defined framework for specifying temporal relations among events in text named TimeML~\cite{pustejovsky2003timeml}, which defines 14 temporal relations (known as \textit{TLINK}) in their latest specification v1.2.1~\cite{timemlSpec}, form where we only use five relations: (a) before, (b) after, (c) immediately before, (f) immediately after, and (e) simultaneous. The relations (a) - (d) reflect two events happening one after another - which we refer to as the BEFORE relation in this work. However, relation (e) can reflect two different cases. First, two actions happen at the same time and the actions' implementation overlaps. E.g., action B and C in Fig.~\ref{fig:example}. Second, two actions happen at the same time but independently. E.g., action D, E, and F in Fig.~\ref{fig:example}. We refer to cases (e1) and (e2) as SIMULTANEOUS-OVERLAP and CONCURRENT, respectively. Finally, we use a temporal relation named NULL to represent no relation between two actions. We summarize the relations in Table~\ref{tab:temp_relations}. 

\begin{table*}[]
    \centering
    \scriptsize
    \caption{List of temporal relations}
    \label{tab:temp_relations}
    
    \begin{tabular}{p{30mm}p{140mm}}

    \toprule

    \textbf{Relation} & \textbf{Definition, example, and explanation} \\ \midrule

    \multirow{3}{*}{BEFORE} & \textbf{Definition:} two techniques happen one after another sequentially as time progresses. \\ 
    {} & \textbf{Example:} \textit{Upon opening the file, the \textbf{user was prompted to enable macros} to complete the form, which began execution of the malware. Once executed, the macro \textbf{created a VBS script} (Updater.vbs), two PowerShell scripts (temp.ps1 and Script.ps1), and installed persistence through a scheduled task.} \\ 
    {} & \textbf{Explanation:} Here, \textit{T1059: Command and Script Interpreter} technique is performed after \textit{T1204: User Execution}. Hence, $(T1204, T1059)$ have a \textit{BEFORE} relation. \\ \midrule

    \multirow{3}{*}{SIMULTANEOUS-OVERLAP} & \textbf{Definition:} two techniques are executed together and one requires the other for execution \\

    {} & \textbf{Example:} \textit{A \textbf{scheduled task} was then created to assist in execution of the \textbf{keylogger}.} \\ 

    {} & \textbf{Explanation:} Here, \textit{T1053: Scheduled Task} technique is performed to facilitate the execution of \textit{T1056: Input Capture}. Hence, $(T1053, T1056)$ have an \textit{SIMULTANEOUS-OVERLAP} relation. \\ \midrule
    
    \multirow{3}{*}{CONCURRENT} & \textbf{Definition:} two techniques executed independently, however, happen at the same phase of an attack. \\
    {} & \textbf{Example:} \textit{all of which were executed via PowerShell cmdlets or built-in Windows utilities like \textbf{whoami}, net, \textbf{time}, tzutil and tracert;} \\ 
    {} & \textbf{Explanation:} Here, \textit{T1033: System User/Owner Discovery}, and \textit{T1124: System Time Discovery} are executed independently, but they happen during the discovery phase of an attack. Hence, $(T1033, T1124)$ have an \textit{CONCURRENT} relation.  \\ \midrule 

    \multirow{3}{*}{NULL} & \textbf{Definition:} No temporal relation exists between two techniques \\
    {} & \textbf{Example:} \textit{all of which were executed via PowerShell cmdlets or built-in Windows utilities like \textbf{whoami}, net, \textbf{time}, tzutil and tracert;} \\ 
    {} & \textbf{Explanation:} This example does not talk about \textit{T1566: Phishing}, and \textit{T1204: User Execution}. Hence, in the context of this sentence,  $(T1204, T1566)$ has a \textit{NULL} relation. \\

    \bottomrule
    \end{tabular}
    
\end{table*}

\textbf{Temporal Attack Pattern:} Two attack actions with the same temporal relation may exist in multiple cyberattacks, indicating a pattern of adversary behavior in performing a chain of malicious actions. We refer to the pattern as a temporal attack pattern - the frequent occurrence of two techniques with a temporal relation. E.g.:, in Example 1, two different cyberattacks in excerpt 1 and excerpt 2 contain the same temporal relation between two same techniques: phishing is followed by macro execution - which we can refer to as \textit{phishing} BEFORE \textit{macro execution}. In this work, we first identify temporal relations among actions and then identify the patterns by finding recurring relations among the same pair of techniques with the same temporal relation.  

\section{Methodology}
\label{sec:method}
\begin{figure}
    \centering
    \includegraphics[width=\columnwidth]{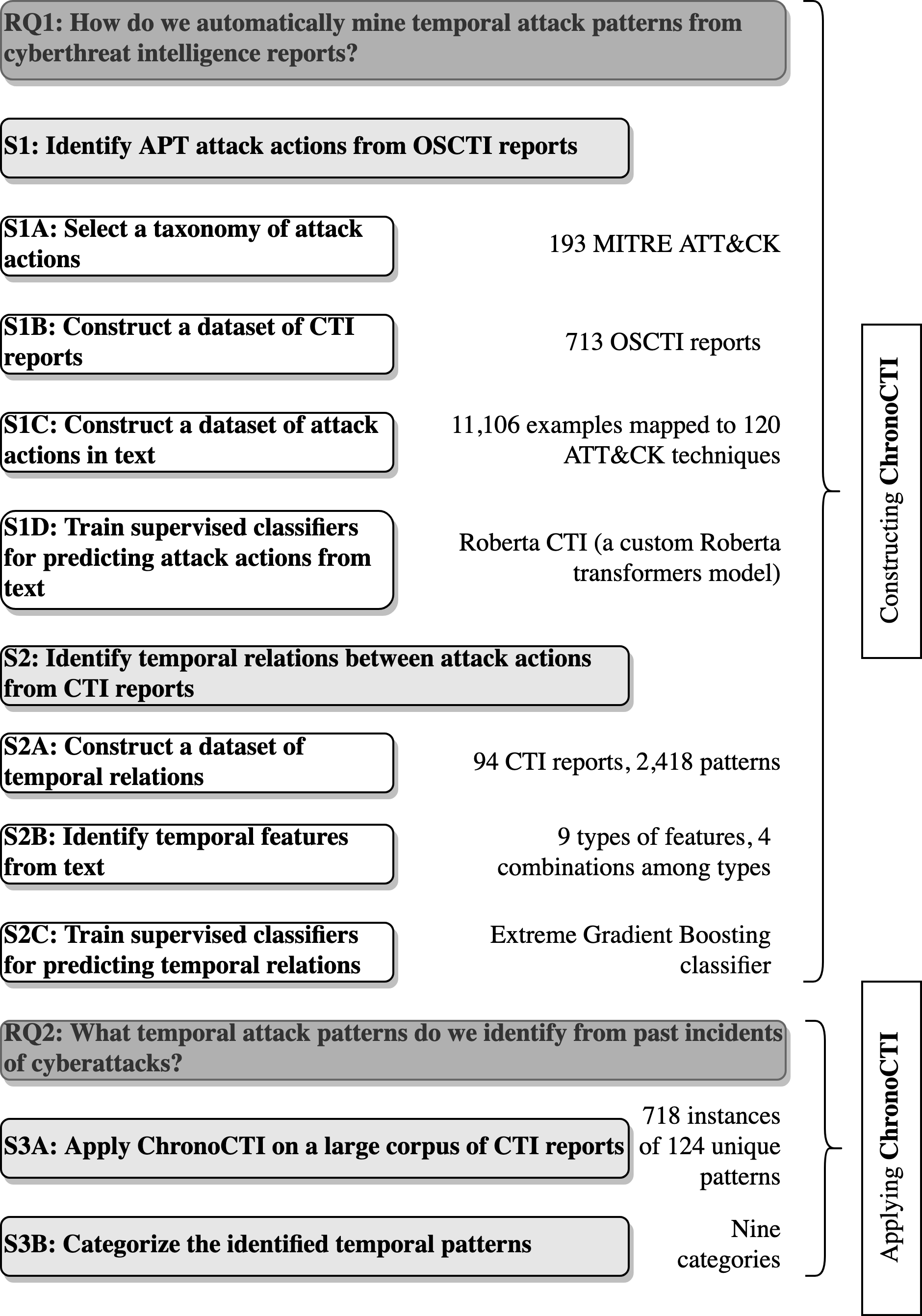}
    \caption{Methodology for RQ1 and RQ2}
    \label{fig:method}
\end{figure}

We discuss the methodology for RQ1 and RQ2 in this section. We construct \textbf{ChronoCTI} in RQ1. We apply \textbf{ChronoCTI} on a large corpus of CTI reports in RQ2. We show the high-level overview of the methodology, along with the corresponding output from each step in Fig.~\ref{fig:method}. 

\noindent \fbox{\begin{minipage}{26em}
RQ1 consists of two steps. The first step is \textbf{S1: identify attack actions from CTI reports} - where we classify sentences into actions. The second step is \textbf{S2: identify temporal relations between attack actions from CTI reports} - where we classify the temporal relations between actions identified from S1. We perform the sub-steps of \textbf{S1} and \textbf{S2} sequentially, as described in the following subsections. 
\end{minipage}} 

\textbf{S1A: Select a taxonomy of attack actions:} We first choose a taxonomy for known adversary actions. Several such taxonomies exist: (a) MITRE ATT\&CK, (b) Cyber-kill-chain~\cite{killchain}, (c) VERIS~\cite{veris}, and (d) CAPEC~\cite{capec}. We choose MITRE ATT\&CK as the taxonomy for the following reasons. MITRE ATT\&CK is also well known to cybersecurity practitioners, and MITRE actively collaborates with practitioners to update its taxonomy and repository~\cite{strom2018mitre}. MITRE ATT\&CK is also prevalently used to identify attack actions in online CTI reports. Moreover, the cyber-kill chain focuses only on high-level actions, which is too abstract for our task. VERIS is a relatively newer framework, and CAPEC is more tailored to application security weaknesses.


\textbf{S1B: Construct a dataset of CTI reports:} To mine temporal patterns from CTI reports, we first need a dataset of CTI reports where we train and validate \textbf{ChronoCTI}. We use the official website of MITRE ATT\&CK as the single source for collecting CTI reports for the following reasons. As the MITRE ATT\&CK repository is maintained through the collaboration of practitioners from multiple security vendors, any citation used in MITRE ATT\&CK indicates that the corresponding CTI reports are of acceptable quality and sufficient technical description. However, alternative options exist to collect CTI reports, which we did not exercise for the following reasons. The first option is to go to the websites of reputed security vendors and find articles on attacks. However, the option is not feasible because the vendor websites contain attack-relevant and irrelevant articles that need to be manually inspected. The second option is to use a well-known curated repository of CTI reports in pdf format called \textit{APTNotes}~\cite{githubaptnote}. We inspected several reports from the repository and found that the reports primarily contain technical descriptions of multiple cyberattacks in a single document, along with contents irrelevant to attacks, such as executive summary, attack trends, details of attack impacts, and countermeasures. Filtering irrelevant portions of the text from the irrelevant ones and separating multiple attacks from one another requires a significant manual inspection. 


\textit{Filtering criteria:} Our purpose is to mine temporal patterns from reports, and hence, we only need reports that describe an attack incident written as a story with a sequentially ordered event narrative. Hence, we establish inclusion and exclusion criteria for filtering the citations so that we only have citations for CTI reports that describe cyberattack incidents in temporal order. \textbf{Inclusion criteria}: Content of a citation URL: (a) must be in English; (b) must be accessible online; (c) describes cyberattack incidents step by step as time progresses. \textbf{Exclusion criteria} Content of a citation URL: (a) is detected by the browser as insecure; (b) does not contain cyberattack description; and (c) is a social media post, definition of a cybersecurity concept (e.g., Wikipedia, NIST glossary), API documentations, source code repositories, (d) describes malware features or cybercriminals capabilities, (e) does not describe an attack in a time-progression order, (f) non-attack relevant content such as countermeasures, security tools, statistics, impact statement, generic security news, and (g) is a pdf, presentation slides or audio/video media documents. Although PDF documents may be a good source for describing attacks in temporal order, we discarded them for similar reasons to not select \textit{APTNotes}. Moreover, we also found that PDF documents are often a periodic compilation of multiple attacks published as web articles in the past, which we would have already obtained from the inclusion criteria. 

At the end of applying the criteria, we have a set of CTI reports that contain attack descriptions as a sequence of events as time progresses. We refer to the set of reports as \textit{curated CTI report set}. We need a dataset to train a learner to predict temporal relations. To our knowledge, no such dataset exists, and we need to construct the dataset from scratch. However, this task is challenging for the following reasons. First, one must be sufficiently familiar with ATT\&CK techniques to build such a dataset. Second, one needs to identify one or more techniques from each sentence in the report and then identify their temporal relations by manual investigation. Third, identifying the temporal order of events is also a subjective task. Fourth, reports are from different security vendors and authors. Understanding the temporal patterns may be straightforward or difficult depending on the writing style. Overall, these challenges introduce cognitive biases while we build the dataset. 

\textit{Construct an evaluation set of CTI reports:} We investigated the \textit{curated CTI report set} and found that a portion of the reports mentions the used ATT\&CK techniques at the end. We refer to the portion as \textit{Set A}. Moreover, we also found a set of reports from a specific security vendor named \textit{The DFIR Reports}\footnote{https://thedfirreport.com/}. The vendor's reports included the observed techniques and provided a graphical timeline of events, which is very helpful for building the dataset. We visited their website and included all the CTI reports if they provided a list of techniques at the end of the report and also fulfilled all the inclusion and exclusion criteria. We refer to newly added reports from THE DFIR REPORT as \textit{Set B}. We then combine \textit{Set A} and \textit{Set B} - which we refer to as the \textit{curated CTI report set for evaluation}. We use the reports from the set for building the dataset because the list of ATT\&CK techniques provided at the end helps us identify the techniques and keeps us from missing any techniques.

\textbf{S1C: Construct a dataset of attack actions in text:} As CTI reports are written in English, the sentences describe the attack action. We first need to predict attack action. We formulate the prediction as a text classification task, where a machine learning model can predict the corresponding ATT\&CK techniques given a piece of text. For example, consider the following sentence, \textit{that those documents were likely delivered through spear-phishing attachments in emails}~\cite{dfirReportIcedIdMalware}. We need to build a classifier that can predict the described ATT\&CK technique in the sentence - in this specific case, the technique is \textit{T1566: Phishing}. To train the text classifier, we obtain the procedure enumeration from the definitions of ATT\&CK techniques. We use the procedure examples as training sentences and ATT\&CK techniques as their corresponding labels. We use the procedure examples of sub-techniques as examples for their parent technique. 
We observed the following challenges in this prediction task. First, a sentence may not describe any technique at all. For example, \textit{All of these factors align to point to the [REDACTED COUNTRY NAME] Oilrig group as the likely threat actors behind this intrusion}~\cite{dfirReportCollectExfiltrateSleepRepeat}. Second, a sentence can describe more than one technique. For example, \textit{A scheduled task was then created to assist in execution of the keylogger}~\cite{dfirReportCollectExfiltrateSleepRepeat}. Here, we identify two techniques: \textit{T1053: Scheduled Task/Job}, and \textit{T1056: Input Capture}. However, procedure examples in ATT\&CK are a one-to-one mapping between a sentence and a technique. Third, to train a text classifier, we must provide sufficient examples. However, ATT\&CK lists 193 techniques, but not all have sufficient examples. To address the first two challenges, we formulate the prediction task as a multi-class, multi-label text classification task. We only keep techniques with at least 20 examples to address the third challenge. 

We conduct a round of mapping sentences to techniques to address the issue of missing mappings obtained from ATT\&CK. We apply c-TFIDF~\cite{ctfidf} as a weak classifier to get a prediction of one or more techniques given a sentence. The first author reviews the prediction made by c-TFIDF. Then, the fourth author reviews the prediction made by c-TFIDF of a set of randomly sampled 15\% of the procedure examples. Then, the authors resolve their disagreements. We also keep the original mapping as provided in ATT\&CK. Thus, we obtain a multi-class, multi-label dataset. 

\textbf{S1D: Train supervised classifiers for predicting attack actions from text:} We must represent text as a vector of features to train a learner. The following language models can be candidates: TFIDF, word embeddings, and large language models. However, we discard TFIDF because the model cannot compute features from the out-of-vocabulary words. We use word embeddings, transformers-based large language models, and text generation-based large language models as candidates. We describe the models in Table~\ref{tab:language_models}. We represent the procedure examples using the stated language models, and then we train a multi-class, multi-label text classifier. For \textit{Spacy-lg}~\footnote{https://spacy.io/models/en\#en\_core\_web\_lg}, \textit{CTI-W2V}, \textit{Roberta}~\footnote{https://huggingface.co/roberta-base}, and \textit{CTI-Roberta}, we use a deep neural network with transformers architecture provided from \textit{Spacy}. Finally, for fine-tuning the \textit{GPT3.5 (gpt-3.5-turbo-1106)~\footnote{https://platform.openai.com/docs/guides/fine-tuning}}, we use the deep neural network architecture provided by OpenAI, which is hosted on the OpenAI platform. We then compare the performance of the language model and classifier combinations. We use the following metric for comparison: (a) macro F1, which is the average F1 score for all labels~\cite{opitz2019macro}, and (b) binary cross entropy loss~\cite{rubinstein2004cross}. We use the best model for the next steps.

\noindent \fbox{\begin{minipage}{26em}
\textbf{End of S1, and beginning of S2}: At this point, we can predict attack actions from CTI reports, but not temporal relations. We need another learner to predict the relations. We perform the following steps. 
\end{minipage}} 

\textbf{S2A: Construct a dataset of temporal relations:} We define the task of predicting the temporal relations of adversary actions as the following. Let a CTI report $R$ may contain one or multiple techniques from all the ATT\&CK techniques: $T_1, T_2, ..., T_n$. We then construct $(n \times n)$ pairs of techniques. If, in $R$, we observe that the two techniques in the pair are executed with respect to time progression, then the pair has a temporal relation between them. Our task is to train a model that takes $R$ as an input and predicts one or more temporal relations among all possible pairs of techniques. The first three authors manually investigate all the reports from \textit{curated CTI report set for evaluation} and identify temporal relations. After that, the authors resolve their disagreements through negotiated agreement~\cite{negotiatedagreement}. 

\begin{table*}[]
    \centering
    \scriptsize
    \caption{List of language models we use to classify a piece of text to attack action}
    \label{tab:language_models}
    
    \begin{tabular}{llp{145mm}}
    \toprule
    \textbf{Model} & \textbf{Type} & \textbf{Description} \\ \midrule

    Spacy-lg & Default & Spacy is a well-known Python package for natural language processing practitioners. Spacy provides pre-trained word embedding named \textit{Spacy-lg}, which is trained on Ontonotes and large web corpora by Spacy. \\

    CTI-W2V & Custom* & A custom word embedding model on three thousand web articles found from citations used in MITRE ATT\&CK. The citations include 1,301 citations from technique definitions and other miscellaneous citations. We use \textit{gensim}~\footnote{https://github.com/piskvorky/gensim} package for the task and used default hyper-parameters. \\

    Roberta & Default & The state-of-the-art transformer-based large language model from BERT, trained on large web corpora by Google. We obtain the model from the Huggingface repository. \\

    CTI-Roberta & Custom* & A fine-tuned Roberta model on three thousand articles, as mentioned in CTI-W2V. We use \textit{simpletransformer}~\footnote{https://github.com/ThilinaRajapakse/simpletransformers} package and fine-tuned the model with default hyper-parameters, with batch sizes of 8, 16, and 32. We then select the model with the lowest cross-entropy loss from the three models trained from three batch sizes. \\

    OpenAI & Custom* & A fine-tuned GPT3.5 text generation model (\textit{gpt-3.5-turbo-1106}) for predicting sentence techniques. The model is hosted on their own systems, and we have limited control over the training architecture, except for a few hyperparameters. \\
    \bottomrule
    \multicolumn{3}{c}{*the model is further fine-tuned on cybersecurity-specific text}
    \end{tabular}
\end{table*}

\textbf{S2B: Identify temporal features from text}
\label{sec:featureset}
We need to identify features to train a learner on the features to predict temporal relations among pairs of techniques. We describe the features below. The features are numeric attributes of a pair of techniques for a CTI report $R$: $(T_x, T_y)$.

\textit{Default feature: Probability of $T_x$, and $T_y$ in sentences of $R$} 
If there is a relation between $T_x$ and $T_y$, then these two techniques should be present in the report, and the classifier from step S1 would predict the techniques with higher probability. If $R$ contains $m$ sentences, we calculate the probabilities of $T_x$ and $T_y$ for $m$ sentences, and then we take the top five probabilities for $T_x$ and $T_y$ as features. 

\begin{table}[]
    \centering
    \scriptsize
    \caption{List of temporal markers}
    \label{tab:temporalMarkers}
    \begin{tabular}{lp{45mm}}
         \toprule
         \textbf{Relation} & \textbf{Temporal marker} \\ \midrule
         BEFORE & after, afterward, following, immediately, instantly, later, next, then, succeeding, subsequent, subsequently, before, previous, prior, previously, preceding \\ 

         SIMULTANEOUS-OVERLAP & during, while, within, through, throughout \\ 

         CONCURRENT & concurrent, concurrently, contemporary, simultaneous, simultaneously \\

         \bottomrule
    \end{tabular}
\end{table}

\textit{F1: Time signal heuristics}
\begin{enumerate*}[leftmargin=6mm, label=(\alph*)]
    \item \textbf{Temporal markers in text:} Natural language often contains specific words that can denote the presence of temporal relations, such as \textit{before}, \textit{after}. We obtain a set of temporal markers from~\cite{derczynski2017automatically} and use them as heuristics as shown in Table~\ref{tab:temporalMarkers}.
    \item \textbf{TimeML features:} \textit{TimeML} is a specification to specify temporal event sequences in text. For example, the sentence \textit{Upon opening the file, the user was prompted to enable macros}~\cite{dfirReportCollectExfiltrateSleepRepeat} can be specified using \textit{TimeML} as following: \verb|e1=open, e2=enable, rel=before|. We use a tool called \textit{Tarsqi-ttk}~\footnote{http://timeml.org/site/tarsqi/toolkit/} to get the TimeML annotations of events from a report. For each event identified by Tarsqi-ttk, we compute the parent noun or verb phrases and then find the corresponding techniques. We then record the event relation types, as defined in TimeML. We then aggregate the TimeML relation types per pair of techniques and record counts of relations seen for each pair of techniques. 
\end{enumerate*}

\textit{F2: Sentence level feature}
\begin{enumerate*}[leftmargin=6mm, label=(\alph*)]
    \item \textbf{Sentence adjacency:} If $T_x$ and $T_y$ are related temporarily, the corresponding sentences should be closely located. We count the occurrences where two sentences, with a distance $d = -4, ..., 0, ..., 4$, contain $T_x$ and $T_y$, respectively. $d=0$ for two sentences next to each other. $d=1$ means another sentence separates two sentences. $d=-1$ means $T_y$ is in the first sentence, and $T_x$ is in the second sentence. 

    \item \textbf{Same sentence:} If $T_x$ and $T_y$ are related temporarily, they may share the same sentence. We compute the count of sentences where $T_x$ and $T_y$ are found. 

    \item \textbf{Sentence similarity:} If $T_x$ and $T_y$ are temporarily related, the sentences containing the two techniques may have a higher semantic similarity. Using word embeddings, we compute the cosine similarity~\cite{howarth2017dictionary} between sentences where $T_x$ is present and where $T_y$ is present. For the embedding, we use the CTI-W2V word embedding as proposed in Step S1D. 

    \item \textbf{Coreference:} If $T_x$ and $T_y$ are related temporarily, then the sentences containing the two techniques may have a coreference relation~\cite{jurafsky2023speech}. E.g., \textit{Upon opening the file, the user was prompted to enable macros to complete the form, which began executing the malware. Once executed, the macro created a VBS script}~\cite{dfirReportCollectExfiltrateSleepRepeat}. Here, \textit{the macro} is coreferencing the \textit{macros} stated in the first sentence. We use \textit{neuralcoref}~\footnote{https://github.com/huggingface/neuralcoref} deep learning model to identify the sentence's coreference relations. 
\end{enumerate*}

\textit{F3: Discourse relations} 
\label{sec:discourse}
Discourse relations indicate the rhetorical structure among the sentences in a paragraph or article~\cite{jurafsky2023speech}. We use five discourse relations between two sentences: S1, and S2 from~\cite{jurafsky2023speech}. 
\begin{enumerate*}[leftmargin=6mm, label=(\alph*)]
    \item \textbf{Next:} S2 describes something that happened after the events described in S1. E.g., \textit{Upon opening the file, the user was prompted to enable macros to complete the form, which began execution of the malware. Once executed, the macro created a VBS script (Updater.vbs), two PowerShell scripts (temp.ps1 and Script.ps1), and installed persistence through a scheduled task}~\cite{dfirReportCollectExfiltrateSleepRepeat}.

    \item \textbf{Elaboration:} S2 explains the action stated in S1. E.g., \textit{The intrusion began with the execution of a malicious macro within a Word document. The document was themed as a job application for the firm Lumen}~\cite{dfirReportCollectExfiltrateSleepRepeat}.

    \item \textbf{If-else:} S2 describes a conditional outcome of an action described in S1. E.g., \textit{if the current user has high privileges, install.bat is executed directly, Otherwise, mshlpweb.dll is executed using rundll32.exe}~\cite{mediumKonni2019CampaignR3}.

    \item \textbf{List:} S1 and S2 have similar semantic meanings - however, contain independent neucleus~\footnote{neucleus refers to the protagonist or antagonist in a sentence}. E.g., \textit{Install.bat — acts as an installer to ensure persistence and execute mshlpsrvc.dll, mshlpweb.dll — acts as loader; responsible to elevate privileges}~\cite{mediumKonni2019CampaignR3}. 

    \item \textbf{Miscellaneous:} S1 and S2 do not have any discourse relation at all, or they have a discourse relation other than Next, Elaboration, If-else, and List. E.g., \textit{Constructs and executes the command-line to download additional files, The document contains three hidden text boxes}~\cite{mediumKonni2019CampaignR3}. 
\end{enumerate*}

We train a discourse relation classifier, which we model as a sentence-pair classification task. We create a dataset of sentence pairs and their corresponding discourse relations. The pairs came from (a) the sentences located one after another in CTI reports and (b) the sentences having coreferences. The first and the third author independently annotate the dataset. Then, the authors resolve their disagreements. On the constructed dataset, we train the classifier using a deep neural network with a transformer architecture based on the CTI-Roberta language model proposed in Step S1D. We use the \textit{simpletransformer} python package for the task. 

\textit{F4: Apriori features} 
We utilize the intuition that if $T_x$ and $T_y$ are related temporarily, they were seen together in past cyberattacks. We can obtain such information from MITRE ATT\&CK's~\cite{attack-design} repository of malware\footnote{https://attack.mitre.org/software/}, cybercrime groups\footnote{https://attack.mitre.org/groups/}, and past campaigns\footnote{https://attack.mitre.org/campaigns/}. We obtain the information and obtain apriori features by performing association rule mining~\cite{piatetsky1991discovery}. We obtain numeric values of the following well-known association rule mining features: (a) support, (b) pointwise mutual information, (c) phi correlation coefficient, (d) causal support, (e) Jaccard distance, (f) confidence, (g) causal confidence, (h) conviction, and (i) added value~\cite{interstingMeasures}. We also performed one-hot categorical embedding~\cite{harris2015digital} by representing the numeric values of apriori features associated with each pair of techniques $(T_x, T_y)$. 

\begin{table}
    \centering
    \scriptsize
    \caption{List of classifiers used as temporal learners}
    \label{tab:temporalClassifiers}
    \begin{tabular}{lp{60mm}}
         \toprule
         \textbf{Type} & \textbf{Classifiers} \\ \midrule
         Ensemble & Adaboost, Extra Trees, Gradient Boosting, Random Forest, Hist Gradient Boosting \\ 

         Linear & Passive Aggressive, Ridge, Stochastic Gradient Descent \\ 

         Naive Bayes & Bernouli, Gaussian, MultiNomial \\

         Neighbors & K-nearest Neighbors, Nearest Centroid \\ 

         Tree & Decision Tree, Extra Tree \\ 
         \bottomrule
    \end{tabular}
\end{table}

\begin{table}[tbh]
\centering
\scriptsize
\setlength{\tabcolsep}{3pt}
\caption{Comparison of language models}
\label{tab:comparison_language_model}
\begin{tabular}{@{}lcccccc@{}}
\toprule
\multirow{2}{*}{\textbf{Language}} & \multicolumn{3}{c}{\textbf{Macro F1}} & \multicolumn{3}{c}{\textbf{Loss}} \\
\cmidrule{2-4} \cmidrule{5-7}
{} & \textbf{Mean} & \textbf{Med.} & \textbf{Sd.} & \textbf{Mean} & \textbf{Med.} & \textbf{Sd.} \\ \midrule
\textbf{Spacy-lg}     & 0.68          & 0.67            & 0.01 & 0.17 & 0.15 & 0.05 \\
\textbf{CTI-w2v}      & 0.66          & 0.66            & 0.01 & 0.20 & 0.21 & 0.34 \\
\textbf{Roberta Base} & 0.75          & 0.76            & 0.01 & 0.45 & 0.44 & 0.06 \\
\textbf{Roberta CTI}  & \textbf{0.77}          & \textbf{0.77}            & 0.01 & 0.41 & 0.41 & 0.10 \\
\textbf{OpenAI GPT-3.5-1106}   & 0.56          & 0.56            & 0.01 & 0.20 & 0.19 & 0.06 \\ \bottomrule
\end{tabular}
\end{table}

\textbf{S2C: Train supervised classifiers for predicting temporal relations}
\label{sec:temporal_learners_list}
We train three types of supervised classifiers on temporal features (default and F1-F4).
\begin{enumerate*}[leftmargin=4mm, label=(\alph*)]
    \item \textbf{Traditional classifiers} primarily refers to the non-deep neural network-based classifiers such as Naive Bayesian, K nearest neighbors, or random first classifiers. We use the following classifiers from the \textit{scikit-learn} packages\footnote{https://scikit-learn.org/stable/modules/classes.html}, as shown in Table~\ref{tab:temporalClassifiers}. 

    \item \textbf{Deep Neural Network based classifiers} Traditionally, deep classifiers do not perform well for small number of features~\cite{arik2021tabnet}, reflecting our case. Hence, we use a deep neural network classifier proposed by Google named \textit{TabNet}~\cite{arik2021tabnet}, which includes both the functionalities of transformers and the attention mechanism. We use \textit{pytorch} package\footnote{https://pytorch.org/} for the task. We also use the multi-layer perception (MLP) classifier from \textit{scikit-learn}.

    \item \textbf{Graph Neural Net Classifiers} The temporal relation prediction task can also be formulated as a graph where techniques are nodes, pairs of techniques are edges, and all the identified features are numeric edge attributes. Then, the task can be formulated as an edge label prediction problem. We use the relational graph convolutional neural network (RGCN) classifier as a predictor~\cite{schlichtkrull2018modeling}, and we use \textit{Deep graph library}\footnote{https://www.dgl.ai/} for performing the task. 
\end{enumerate*}
We evaluate the classifiers on the following metrics\footnote{https://scikit-learn.org/stable/modules/classes.html\#classification-metrics}: precision (P), recall (F), and F1 score (F). However, as the classifiers predict one or more temporal relations for a pair of techniques, we use the macro average of P, R, and F - which are the averages of P, R, and F of the four temporal relation types. We also use three more metrics specially designed for multi-label classification\footnote{https://scikit-learn.org/stable/modules/classes.html\#multilabel-ranking-metrics}: (a) precision @ k (P@K), (b) label ranking average precision (LRAP), and (c) normalized discounted cumulative gain (NDCG). These three metrics denote how well the model can rank the relevant patterns before the irrelevant items based on their prediction probability.



\noindent \fbox{\begin{minipage}{26em}
At this point, we can predict temporal relations among actions from CTI reports - which marks the completion of constructing the pipeline of \textbf{ChrnoCTI}. We will now apply \textbf{ChronoCTI} to identify temporal attack patterns. 
\end{minipage}} 

\textbf{S3A: Apply ChronoCTI on a large corpus of CTI reports}
\label{sec:rq2}
We apply \textbf{ChrnoCTI} on the \textit{curated CTI report set}. If we find a particular temporal relation $\tau$ between a pair of techniques $(T_x, T_y)$ in at least $n$ report, then we refer to the tuple: $(T_x, T_y, \tau)$ as a temporal attack pattern. 

\textbf{S3B: Categorize the identified temporal patterns}
The first two authors categorize the identified patterns orthogonally using open coding~\cite{saldana_coding_2015}. Then the authors resolve their disagreement through negotiated agreement~\cite{negotiatedagreement}. 

\section{Findings on RQ1}
\label{sec:rq1Findings}
This section describes the findings from each step of the methodology of RQ1 (see Section~\ref{sec:method}). We show the overall findings of each step in Fig.~\ref{fig:method}. 

\textbf{S1A: Select a taxonomy of attack actions:} We use version 12.1\footnote{https://github.com/mitre/cti/releases/tag/ATT\%26CK-v12.1} of MITRE ATT\&CK containing 193 techniques. 

\textbf{S1B: Construct a dataset of CTI reports:} We obtain a list of 1,301 citation URLs from the techniques listed in MITRE ATT\&CK. After applying the inclusion and exclusion criteria, we identified 673 CTI reports, which we refer to as \textit{curated CTI report set}. From the \textit{curated CTI report set}, we identify 54 CTI reports, which include the MITRE ATT\&CK techniques observed in the corresponding cyberattacks described in the report. We found an additional 40 reports from THE DFIR Report vendor. We construct the \textit{curated set of CTI reports for evaluation} by combining these 54 and 40 CTI reports, resulting in 94 reports consisting of 7,052 sentences. 

    
    
    

\textbf{S1C: Construct a dataset of attack actions in text:} We obtain a set of 11,106 one-to-one mappings between sentences and techniques from MITRE ATT\&CK. After reviewing the prediction made by c-TFIDF, the two authors obtained a median Cohen's kappa ($\kappa$) score of 0.83 between the first author and the fourth author on a set of randomly sampled 15\%  of the dataset. According to ~\cite{mchugh_interrater_2012}, the score indicates a substantial agreement. We identify 120 techniques from the dataset with at least 20 examples. 

\textbf{S1D: Train supervised classifiers for predicting attack actions from text:} We train the CTI-Roberta model from the Roberta base model. We identified the lowest loss in training the Roberta CTI with a batch size of 16. For CTI-W2V, Spacy-lg, Roberta Base, and Roberta CTI, we use Spacy's default deep neural network architecture\footnote{https://spacy.io/usage/processing-pipelines} except for maximum steps and patience - which are set to 5,000 and 750, respectively. For fine-tuning the OpenAI LLM,  We use the default hyper-parameters. In Table~\ref{tab:comparison_language_model}, we compare the five models after performing five-fold cross-validation with a split of 80\% training and 20\% testing set. CTI-Roberta model performs best and slightly outperforms the Roberta base model in classifying a sentence into multiple techniques. Moreover, the loss scores of the Roberta base and CTI-Roberta model suggest that if we train these two models in further epochs, we will gain even better performance from those two models. The worst-performing model in our experiment is the GPT3.5 LLM model from OpenAI. However, this specific performance of the GPT-3.5 is observed at the default hyper-parameter, with an average cost of USD 5 per fold. The loss score of the GPT-3.5 model suggests that further fine-tuning could lead to a slightly better performance, with the expense of additional financial costs. However, the loss scores of GPT-3.5, Spacy-lg, and CTI-w2v models are almost half of the loss score of the two Roberta models - which indicates that, with unlimited training, Roberta model could lead to a better performance than that of the rest of the models. \textbf{Hence, we choose CTI-Roberta as the language model for the next steps.}

We also evaluate the performance of the Roberta CTI model on a set of unseen CTI reports, which we get from \textit{curated CTI report set for evaluation}. We use this set because the reports in this set mention the techniques used in the corresponding cyberattack. We evaluate step by step as follows. First, Each report contains multiple sentences. We assume that the classifier predicts a technique in a report if the classifier predicts the technique in at least one sentence from the report. Second, While a classifier makes a prediction, the classifier predicts the presence of a technique with a probability. We evaluate the classifier's performance across prediction probabilities as $[0.05, 0.10, 0.15, ..., 0.95]$. Finally, we use precision (P), recall (R), and F1 score (F) as the evaluation metric. We observe that the precision and recall converge at the prediction probability of 0.95 - where the highest F1 score (0.69) is observed at this probability. \textbf{In the following steps, we use the prediction from the CTI-Roberta model at the 0.95 prediction threshold. }


\begin{table}[]
    \centering
    \setlength{\tabcolsep}{3pt}
    \scriptsize
    \caption{Classification performance, and inter-rater agreement for the discourse relation classifier, and the discourse relation dataset respectively}
    \label{tab:discourse}
    \begin{tabular}{llrrrrr}
    \toprule
    \textbf{Discourse relation} & \textbf{Pair type} & \textbf{Count} & P & R & F & \textbf{$\kappa$} score \\ \midrule
    
    \multirow{2}{*}{ELABORATION} & Adjacent & 2087 & 0.77 & 0.87 & 0.82 & \multirow{2}{*}{0.49*} \\
    {} & Coreferenced & 762 & 0.58 & 0.65 & 0.61 & {} \\ \midrule
    
    \multirow{2}{*}{IF\_ELSE} & Adjacent & 81 & 0.89 & 0.89 & 0.89 & \multirow{2}{*}{0.61**} \\
    {} & Coreferenced & 21 & 0.75 & 0.60 & 0.67 & {} \\ \midrule
    
    \multirow{2}{*}{LIST} & Adjacent & 245 & 0.87 & 0.63 & 0.73 & \multirow{2}{*}{0.56*} \\
    {} & Coreferenced & 41 & 1.00 & 0.20 & 0.33 & {} \\ \midrule
    
    \multirow{2}{*}{MISC} & Adjacent & 1258 & 0.68 & 0.74 & 0.71 & \multirow{2}{*}{0.47*} \\
    {} & Coreferenced & 1163 & 0.72 & 0.71 & 0.72 & {} \\ \midrule
    
    \multirow{2}{*}{BEFORE} & Adjacent & 1278 & 0.83 & 0.62 & 0.71 &  \multirow{2}{*}{0.40*} \\
    {} & Coreferenced & 672 & 0.67 & 0.63 & 0.65 & {} \\ \midrule
    
    \multicolumn{3}{c}{*moderate agreement, **substantial agreement~\cite{mchugh_interrater_2012}} \\ \bottomrule
    \end{tabular}
\end{table}

\textbf{S2A: Construct a dataset of temporal relations:} The first three authors independently reviewed the 94 CTI reports in \textit{curated CTI report set for evaluation}. The first author reviews all reports. Four reports were used as examples for the second and third authors. Among the rest, the second author reviews 69, and the third reviews 19 reports. After reviewing, the authors resolve their disagreements through negotiated agreement. The dataset contains 1,342,320 rows, each representing a pair between techniques and the corresponding temporal label. CONCURRENT, BEFORE, and SIMULTANEOUS-OVERLAP have been mapped to 179, 1,208, and 1,031 rows. The rest of the rows are mapped to NULL. The $\kappa$ agreement scores of CONCURRENT, BEFORE, and SIMULTANEOUS-OVERLAP are 0.49, 0.47, and 0.47, respectively - which is moderate agreement~\cite{mchugh_interrater_2012}. 


\begin{table}[t]
    \centering
    \scriptsize
    \caption{Performance of Temporal Relation Classifiers}
    \label{tab:temporalClassifierPerformance}
    \begin{tabular}{llrrr}
    \toprule
    \textbf{Classifier}                        & \textbf{Features}                                      &   \textbf{P} &   \textbf{R} &   \textbf{F} \\
    \midrule
     \textbf{Gradient Boosting}                  & F1+F2+F3+F4 &        0.87 &     0.61 & \textbf{0.71} \\
     Hist Gradient Boosting & F1+F2+F3+F4 &        0.79 &     0.56 & 0.65 \\
     Decision Tree         & F1+F2+F3+F4 &        0.64 &     0.64 & 0.64 \\
     \textbf{Random Forest}         & F1+F2+F3+F4 &        \textbf{0.9}  &     0.54 & 0.64 \\
     Extra Trees           & F1+F2+F3+F4 &        0.86 &     0.52 & 0.63 \\
     MLP                  & F1+F2+F3+F4 &        0.86 &     0.51 & 0.61 \\
     Stochastic Gradient Descent                  & F1+F2+F3+F4 &        0.87 &     0.52 & 0.61 \\
     RGCN                           & F1+F2+F3+F4 &        0.74 &     0.51 & 0.58 \\
     Passive Aggressive    & F1+F2+F3+F4 &        0.66 &     0.58 & 0.56 \\
     K-nearest Neighbors           & F1+F2+F3+F4 &        0.76 &     0.48 & 0.56 \\
     Adaboost             & F1+F2+F3+F4 &        0.75 &     0.47 & 0.56 \\
     Extra Tree            & F1+F2+F3+F4 &        0.59 &     0.55 & 0.56 \\
     BernoulliNB                    & F1+F2+F3+F4 &        0.48 &     0.66 & 0.51 \\
     BernoulliNB                    & F1+F2                     &        0.48 &     0.65 & 0.51 \\
     BernoulliNB                    & F1+F2+F3           &        0.47 &     0.65 & 0.5  \\
     BernoulliNB                    & F1                                  &        0.49 &     0.62 & 0.5  \\
     RGCN                           & F1+F2                     &        0.64 &     0.41 & 0.48 \\
     MultinomialNB                  & F1+F2+F3+F4 &        0.63 &     0.39 & 0.45 \\
    RGCN                           & F1                                  &        0.67 &     0.4  & 0.45 \\
     RGCN                           & F1+F2+F3           &        0.63 &     0.39 & 0.44 \\
     Extra Trees           & F1+F2+F3           &        0.57 &     0.4  & 0.43 \\
     Decision Tree         & F1+F2+F3           &        0.44 &     0.43 & 0.43 \\
     Decision Tree         & F1+F2                     &        0.4  &     0.42 & 0.41 \\
     Extra Trees           & F1+F2                     &        0.6  &     0.38 & 0.41 \\
     Extra Tree            & F1+F2+F3           &        0.41 &     0.41 & 0.41 \\
     Gradient Boosting                  & F1+F2                     &        0.53 &     0.38 & 0.4  \\
     Extra Tree            & F1+F2                     &        0.4  &     0.41 & 0.4  \\
     Passive Aggressive    & F1+F2                     &        0.41 &     0.39 & 0.4  \\
     Gradient Boosting                  & F1+F2+F3           &        0.5  &     0.38 & 0.39 \\
     Adaboost             & F1+F2                     &        0.46 &     0.37 & 0.39 \\
     Adaboost             & F1+F2+F3           &        0.45 &     0.37 & 0.39 \\
     Hist Gradient Boosting & F1+F2+F3           &        0.52 &     0.38 & 0.39 \\
     Hist Gradient Boosting & F1+F2                     &        0.48 &     0.38 & 0.39 \\
     Gradient Boosting                  & F1                                  &        0.43 &     0.37 & 0.38 \\
     Passive Aggressive    & F1+F2+F3           &        0.42 &     0.37 & 0.38 \\
     RidgeClassifier                & F1+F2+F3+F4 &        0.61 &     0.36 & 0.38 \\
     K-nearest Neighbors           & F1+F2                     &        0.49 &     0.37 & 0.38 \\
     K-nearest Neighbors           & F1                                  &        0.45 &     0.37 & 0.38 \\
     Random Forest         & F1+F2                     &        0.57 &     0.37 & 0.38 \\
     Hist Gradient Boosting & F1                                  &        0.46 &     0.37 & 0.38 \\
     Extra Trees           & F1                                  &        0.43 &     0.37 & 0.38 \\
     Adaboost             & F1                                  &        0.46 &     0.37 & 0.38 \\
     Random Forest         & F1+F2+F3           &        0.57 &     0.37 & 0.38 \\
     TabNet                         & F1+F2+F3+F4 &        0.55 &     0.37 & 0.38 \\
     MLP                  & F1+F2+F3           &        0.39 &     0.36 & 0.37 \\
     TabNet                         & F1                                  &        0.38 &     0.36 & 0.37 \\
     Stochastic Gradient Descent                  & F1+F2+F3           &        0.38 &     0.37 & 0.37 \\
     Stochastic Gradient Descent                  & F1+F2                     &        0.38 &     0.37 & 0.37 \\
     Stochastic Gradient Descent                  & F1                                  &        0.38 &     0.37 & 0.37 \\
     TabNet                         & F1+F2                     &        0.38 &     0.37 & 0.37 \\
     RidgeClassifier                & F1+F2+F3           &        0.51 &     0.37 & 0.37 \\
     RidgeClassifier                & F1+F2                     &        0.51 &     0.37 & 0.37 \\
     RidgeClassifier                & F1                                  &        0.39 &     0.37 & 0.37 \\
     Random Forest         & F1                                  &        0.46 &     0.37 & 0.37 \\
     K-nearest Neighbors           & F1+F2+F3           &        0.49 &     0.37 & 0.37 \\
     Extra Tree            & F1                                  &        0.36 &     0.38 & 0.37 \\
     MLP                  & F1                                  &        0.37 &     0.36 & 0.37 \\
     Nearest Centroid                & F1+F2+F3+F4 &        0.32 &     0.7  & 0.37 \\
     MLP                  & F1+F2                     &        0.38 &     0.37 & 0.37 \\
     TabNet                         & F1+F2+F3           &        0.37 &     0.35 & 0.36 \\
     Decision Tree         & F1                                  &        0.35 &     0.38 & 0.36 \\
     GaussianNB                     & F1+F2                     &        0.31 &     0.87 & 0.35 \\
     Nearest Centroid                & F1                                  &        0.31 &     0.61 & 0.35 \\
     Nearest Centroid                & F1+F2                     &        0.32 &     0.61 & 0.35 \\
     Nearest Centroid                & F1+F2+F3           &        0.32 &     0.61 & 0.35 \\
     GaussianNB                     & F1                                  &        0.31 &     0.84 & 0.34 \\
     \textbf{GaussianNB}                     & F1+F2+F3           &        0.28 &     \textbf{0.89} & 0.31 \\
     MultinomialNB                  & F1+F2+F3           &        0.35 &     0.28 & 0.28 \\
     MultinomialNB                  & F1+F2                     &        0.35 &     0.27 & 0.27 \\
     MultinomialNB                  & F1                                  &        0.34 &     0.25 & 0.25 \\
     GaussianNB                     & F1+F2+F3+F4 &        0.28 &     0.81 & 0.25 \\
     Passive Aggressive    & F1                                  &        0.31 &     0.25 & 0.24 \\ \bottomrule
    \end{tabular}
\end{table}

\textbf{S2B: Identify temporal features from text:} We discussed five types of feature sets in Section~\ref{sec:featureset}, which results in 309 numeric features. We report the number of features per feature set as follows: (a) default feature = 10, (b) F1 = 17, (c) F2 = 13, (d) F3 = 10, and (e) F4 = 259. We discussed the methodology for identifying a set of five discourse relations between sentences in Section~\ref{sec:discourse}. The discourse relation dataset contains 7,598 sentence pairs of 4944 adjacent and 2694 coreferenced sentence pairs. These sentence pairs are collected from 73 CTI reports from the \textit{curated CTI report set for evaluation}. We trained two classifiers for discourse identification, one for the adjacent sentence pairs and another for the coreferenced sentence pairs. We show the classifier performance and inter-rater agreement score in Table~\ref{tab:discourse}. 

\textbf{S2C: Train supervised classifiers for predicting temporal relations:} We train the selected classifiers (Table~\ref{tab:temporalClassifiers}) on the \textit{curated CTI report set for evaluation} consisting of 94 reports. Among the 94, 73 reports were used as the training set, and 21 reports were used as the evaluation set. We first performed a five-fold cross-validation of all the selected classifiers across all the feature sets. Note that, for all feature combinations, we used the default feature set with additional combinations of features. 

Table~\ref{tab:temporalClassifierPerformance} shows the median of the classifiers' macro precision, macro recall, and macro F1 scores across all four feature sets. We observe that the gradient-boosting classifier performs best. We also observe that ensemble-based learners perform better than neural network-based, linear, or nearest neighbor-based models. Specifically, the boosting models perform at least 10\% better than the best-performing deep neural network model. Although deep neural network-based models perform better with complex and large feature sets, literature shows~\cite{arik2021tabnet} that gradient-boosting learners perform the best with the tabular feature type. In our experiment, we observe the same. 

We observe that the apriori features significantly contribute to the classifiers. Without the features, many learners perform well in the recall score. However, the feature helps improve the precision score in all the classifiers. Without the Apriori feature, naive Bayesian classifiers perform the best. We observe the best precision score from the Random forest classifier. The best recall score is observed from the Naive Bayesian classifiers. However, the gradient boosting classifier performed the best neither in precision nor recall. However, the classifiers score best in the case of the F1 score. Overall, with all the features, the top-performing models perform very well in precision but lack recall.

\begin{table}[]
    \centering
    \scriptsize
    \caption{Performance of Extreme Gradient Boosting classifier on the evaluation dataset}
    \label{tab:evaluation}
    \begin{tabular}{lrrrrrr}
    \toprule
    \textbf{Labels} & \textbf{P} & \textbf{R} & \textbf{F} & \textbf{P@50} & \textbf{P@100} & S \\ \midrule
    CONCURRENT & 0.77 & 0.33 & 0.47 & 0.77 & 0.77 & 60 \\
    BEFORE & 0.65 & 0.34 & 0.45 & 0.76 & 0.72 & 300 \\
    NULL & 0.97 & 0.99 & 0.98 & 1.00 & 1.00 & 12208 \\
    SIMULTANEOUS-OV. & 0.61 & 0.17 & 0.27 & 0.72 & 0.61 & 320 \\ \midrule
    Macro Avg. & 0.75 & 0.46 & 0.54 & 0.81 & 0.78 & 12888 \\ \midrule
    LRAP & \multicolumn{6}{c}{0.97} \\
    NDCG & \multicolumn{6}{c}{0.98} \\ 
    \bottomrule
    \end{tabular}
\end{table}

We applied the best-performing model: the gradient boosting classifier on the evaluation dataset. We summarize the performance of the classifier against the held-out dataset in Table~\ref{tab:evaluation}. We observe the following from the table. First, the classifiers perform almost perfectly for the NULL class. However, the classifier shows decent precision for the positive classes but very low recall. The classifier predicts with accuracy, as confirmed by higher precisions - however, the model misses a lot of true examples, as shown by low recall scores. Second, the precision@50 and precision@100 metrics for all the classes show a decent score, which suggests that the top 50 or top 100 predictions made by the classifier are often accurate. Finally, the LRAP and NDCG scores are very close to one - which also suggests that if we rank the predictions by their probability, top predictions will be correct about 97\% of the time. However, from the support column, we also observe that the 99\% majority of the NULL class heavily skews the macro F1, LRAP, and NDCG. 

\section{Findings on RQ2}
\label{sec:rq2Findings}
We ran gradient-boosted classifier on 713 CTI reports from \textit{curated CTI report set}, and set $n=2$. Our proposed model identifies 124 unique temporal patterns of techniques with 718 instances. Among the unique patterns, we identify (a) BEFORE = 84 having 506 occurrences; (b) SIMULTANEOUS-OVERLAP = 25 having 168 occurrences, and (c) CONCURRENT = 15 having 44 occurrences. The first and second authors then grouped the 124 identified patterns into nine orthogonal categories - which we refer to as temporal attack pattern categories. The $\kappa$ agreement score is 0.49, which suggests a moderate agreement between the two authors.  In the following subsections, we discuss the categories. We report the categories in Table~\ref{tab:categoryBaiting} -~\ref{tab:categoryransomware}. The $\tau$ columns in the tables denote the temporal relation types: B = BEFORE, S = SIMULTANEOUS-OVERLAP, and C = CONCURRENT. The C column denotes the count of reports where the relation is found. We also report one example excerpt from a CTI report from where a specific pattern is found.  

\begin{table}[t]
\centering
\scriptsize
\setlength{\tabcolsep}{2pt}
\caption{Baiting towards malicious execution}
\label{tab:categoryBaiting}
\begin{tabular}{p{75mm}cr}
\toprule
\textbf{Pattern}                                                      & \textbf{$\tau$} & \textbf{C}  \\ \midrule
(P1) T1566: Phishing, T1204: User Execution                               & B                 & 201              \\
(P2) T1204: User Execution, T1105: Ingress Tool Transfer                  & B                 & 47               \\
(P3) T1204: User Execution, T1564: Hide Artifacts                         & S                 & 4                \\
(P4) T1204: User Execution, T1082: System Information Discovery           & B                 & 4                \\
(P5) T1204: User Execution, T1053: Scheduled Task/Job                     & B                 & 3                \\
(P6) T1204: User Execution, T1059: Command and Scripting Interpreter      & B                 & 3                \\
(P7) T1204: User Execution, T1482: Domain Trust Discovery                 & B                 & 3                \\
(P8) User Execution, T1547: Boot or Logon Autostart Execution             & B                 & 3                \\
(P9) T1189: Drive-by Compromise, T1059: Command and Scripting Interpreter & B                 & 2                \\
(P10) T1566: Phishing, T1189: Drive-by Compromise                          & B                 & 2                \\
(P11) T1059: Command and Scripting Interpreter, T1564: Hide Artifacts      & S                 & 2                \\
(P12) T1608: Stage Capabilities, T1189: Drive-by Compromise                & B                 & 2                \\
(P13) T1608: Stage Capabilities, T1566: Phishing                           & B                 & 2                \\
(P14) T1203: Exploitation for Client Execution, T1204: User Execution      & S                 & 2                \\
(P15) T1204: User Execution, T1033: System Owner/User Discovery            & B                 & 2                \\
(P16) T1204: User Execution, T1189: Drive-by Compromise                    & B                 & 2                \\ \bottomrule

\multicolumn{3}{p{85mm}}{Example for P1: As with the \textbf{spearphishing email}, the lure documents' content attempts to convince the victim to \textbf{click on another malicious URL} and download a .ZIP file.~\cite{trendMicroEarthVetalaR998}}

\end{tabular}
\end{table}

\textbf{PC-1: Baiting towards malicious execution:} We identify 16 patterns in this category, with 284 occurrences. We report the patterns in Table~\ref{tab:categoryBaiting}. The table shows that the most prevalent attack pattern is executing malicious files or links after phishing (P1), which is observed in 201 reports. Phishing emails often contain document files with hidden executables (P3, P11) or download links (P10, P12, P13, P16). Clicking these executables or downloading the executables leads to further malicious actions. For example, when a victim user opens a Word document, the hidden macro can download victim OS-specific tools (P2), run arbitrary malicious commands (P6, P9), enumerate system information (P4, P7, P15), run task at the background (P5, P8) , or execute exploit scripts (P14). Overall, we observe that the patterns in this category initiate the attack by tricking users into executing seemingly innocuous files or links.

\begin{table}[]
\centering
\scriptsize
\setlength{\tabcolsep}{2pt}
\caption{Bypassing anti-malware in victim system}
\label{tab:categoryBypassing}
\begin{tabular}{p{75mm}cr}
\toprule
\textbf{Patterns}                                                              & \textbf{$\tau$} & \textbf{C} \\ \midrule
(P17) T1027: Obfuscated Files or Information, T1140: Deobfuscate/Decode Files or Information & B & 22 \\
(P18) T1204: User Execution, T1218: System Binary Proxy Execution                    & B          & 11         \\
(P19) T1001: Data Obfuscation, T1071: Application Layer Protocol                     & S          & 6          \\
(P20) T1071: Application Layer Protocol, T1573: Encrypted Channel                    & S          & 6          \\
(P21) T1105: Ingress Tool Transfer, T1620: Reflective Code Loading                   & B          & 5          \\
(P22) T1106: Native API, T1497: Virtualization/Sandbox Evasion                       & S          & 4          \\
(P23) T1059: Command and Scripting Interpreter, T1562: Impair Defenses               & S          & 4          \\
(P24) T1140: Deobfuscate/Decode Files or Information, T1620: Reflective Code Loading & B          & 4          \\
(P25) T1059: Command and Scripting Interpreter, T1098: Account Manipulation          & S          & 3          \\
(P26) T1027: Obfuscated Files or Information, T1059: Command and Scripting Interpreter       & S & 3  \\
(P27) T1059: Command and Scripting Interpreter, T1620: Reflective Code Loading       & B          & 3          \\
(P28) T1570: Lateral Tool Transfer, T1543: Create or Modify System Process           & B          & 3          \\
(P29) T1570: Lateral Tool Transfer, T1569: System Services                           & B          & 2          \\
(P30) T1055: Process Injection, T1482: Domain Trust Discovery                        & B          & 2          \\
(P31) T1055: Process Injection, T1219: Remote Access Software                        & B          & 2          \\
(P32) T1055: Process Injection, T1071: Application Layer Protocol                    & B          & 2          \\
(P33) T1055: Process Injection, T1049: System Network Connections Discovery          & B          & 2          \\
(P34) T1055: Process Injection, T1016: System Network Configuration Discovery        & B          & 2          \\
(P35) T1218: System Binary Proxy Execution, T1548: Abuse Elevation Control Mechanism & B          & 2          \\
(P36) T1136: Create Account, T1098: Account Manipulation                             & B          & 2          \\
(P37) T1204: User Execution, T1055: Process Injection                                & B          & 2          \\
(P38) T1105: Ingress Tool Transfer, T1140: Deobfuscate/Decode Files or Information   & B          & 2          \\
(P39) T1105: Ingress Tool Transfer, T1112: Modify Registry                           & B          & 2          \\
(P40) T1218: System Binary Proxy Execution, T1559: Inter-Process Communication       & S          & 2          \\
(P41) T1105: Ingress Tool Transfer, T1070: Indicator Removal                         & B          & 2          \\
(P42) T1136: Create Account, T1219: Remote Access Software                           & B          & 2          \\
(P43) T1049: System Network Connections Discovery, T1218: System Binary Proxy Execution      & B & 2  \\
(P44) T1047: Windows Management Instrumentation, T1562: Impair Defenses              & S          & 2          \\
(P45) T1021: Remote Services, T1543: Create or Modify System Process                 & B          & 2         \\ \bottomrule

\multicolumn{3}{p{85mm}}{Example for P21: The threat actors used PowerShell to download and execute a new Cobalt Strike PowerShell beacon in memory on the beachhead host.~\cite{dfirReportBumbleBeeDfir11}} \\ 

\end{tabular}
\end{table}

\textbf{PC-2: Bypassing anti-malware in victim system:}
We identify 29 patterns in this category, with 108 occurrences in total. We report the patterns in Table~\ref{tab:categoryBypassing}. The table shows that the attackers perform a variety of actions to circumvent the system defense. Modern operating systems ship with built-in security mechanisms; users often install third-party anti-malware as well. Nonetheless, adversaries bypass these by following. First, adversaries use proxy execution of malicious code by \verb|regsvr32|, \verb|rundll32.exe|, \verb|cmd.exe|, \verb|svchost.exe|, \verb|WMI| program, and native APIs - which anti-malware fails to notice because these programs and APIs are part of the operating systems (P18., P22, P23, P25, P28, P29, P35, P40, P43, P44, P45). Second, attackers inject malicious code into legitimate processes - hence, anti-malware may not be able to detect the malicious code execution residing inside a legitimate process's memory, such as \verb|explorer.exe| (P30, P31, P32, P33, P34, P37). Third, attackers encrypt their communication - hence, their content cannot be analyzed (P17, P19, P20, P26, P38). Fourth, malicious code can be directly unpacked or downloaded to memory instead of the filesystem - so they are skipped from antivirus scans (P21, P24, P27). Other patterns to bypass anti-malware are to manipulating user accounts (P36, P42), changing system configuration (P39), and deleting malicious traces to hinder forensic activities (P41).  

\begin{table}[]
\centering
\scriptsize
\setlength{\tabcolsep}{2pt}
\caption{Profiling target system}
\label{tab:categoryProfile}
\begin{tabular}{p{75mm}cr}
\toprule
\textbf{Pattern}                                                                  & \textbf{$\tau$} & \textbf{C} \\ \midrule
(P46) T1016: System Network Configuration Discovery, T1082: System Information Discovery & C & 12 \\
(P47) T1016: System Network Configuration Discovery, T1059: Command and Scripting Interpreter     & S          & 5           \\
(P48) T1049: System Network Connections Discovery, T1059: Command and Scripting Interpreter       & S          & 5           \\
(P49) T1059: Command and Scripting Interpreter, T1087: Account Discovery                & S             & 4              \\
(P50) T1033: System Owner/User Discovery, T1482: Domain Trust Discovery                 & C             & 4              \\
(P51) T1059: Command and Scripting Interpreter, T1482: Domain Trust Discovery           & S             & 4              \\
(P52) T1059: Command and Scripting Interpreter, T1082: System Information Discovery     & S             & 3              \\
(P53) T1033: System Owner/User Discovery, T1059: Command and Scripting Interpreter      & S             & 3              \\
(P54) T1087: Account Discovery, T1482: Domain Trust Discovery                           & C             & 3              \\
(P55) T1018: Remote System Discovery, T1570: Lateral Tool Transfer                      & B             & 3              \\
(P56) T1016: System Network Configuration Discovery, T1033: System Owner/User Discovery & C             & 3              \\
(P57) T1482: Domain Trust Discovery, T1570: Lateral Tool Transfer                       & B             & 2              \\
(P58) T1059: Command and Scripting Interpreter, T1083: File and Directory Discovery     & S             & 2              \\
(P59) T1057: Process Discovery, T1082: System Information Discovery                     & C             & 2              \\
(P60) T1049: System Network Connections Discovery, T1087: Account Discovery             & C             & 2              \\
(P61) T1482: Domain Trust Discovery, T1068: Exploitation for Privilege Escalation       & B             & 2              \\ 
(P62) T1135: Network Share Discovery, T1482: Domain Trust Discovery                     & C             & 2              \\
(P63) T1135: Network Share Discovery, T1482: Domain Trust Discovery                     & B             & 2              \\
(P64) T1087: Account Discovery, T1570: Lateral Tool Transfer                            & B             & 2              \\
(P65) T1082: System Information Discovery, T1518: Software Discovery                    & C             & 2              \\
(P66) T1033: System Owner/User Discovery, T1082: System Information Discovery           & C             & 2              \\
(P67) T1047: Windows Management Instrumentation, T1518: Software Discovery              & S             & 2              \\
(P68) T1018: Remote System Discovery, T1482: Domain Trust Discovery                     & C             & 2              \\
(P69) T1033: System Owner/User Discovery, T1570: Lateral Tool Transfer                  & B             & 2              \\
(P70) T1018: Remote System Discovery, T1021: Remote Services                            & B             & 2              \\
(P71) T1016: System Network Configuration Discovery, T1518: Software Discovery          & C             & 2              \\
(P72) T1033: System Owner/User Discovery, T1068: Exploitation for Privilege Escalation  & B             & 2              \\
(P73) T1003: OS Credential Dumping, T1087: Account Discovery                            & B             & 2              \\
(P74) T1021: Remote Services, T1057: Process Discovery                                  & B             & 2              \\
(P75) T1016: System Network Configuration Discovery, T1482: Domain Trust Discovery      & C             & 2              \\
(P76) T1016: System Network Configuration Discovery, T1087: Account Discovery           & C             & 2              \\
(P77) T1016: System Network Configuration Discovery, T1570: Lateral Tool Transfer       & B             & 2              \\ \midrule

\multicolumn{3}{p{85mm}}{Example for P46: The threat actors began executing basic discovery commands, all of which were executed via PowerShell cmdlets or built-in Windows utilities like whoami, net, time, tzutil and tracert.~\cite{dfirReportCollectExfiltrateSleepRepeat}}
\end{tabular}
\end{table}

\textbf{PC-3: Profiling target system:}
We identify 32 patterns in this category, with 91 occurrences in total. We report the patterns in Table~\ref{tab:categoryProfile}. The table shows that adversaries collect information related to operating systems, networks, software, and domain accounts at the same phase of time, usually after initial compromises (P46, P50, P54, P56, P59, P60, P62, P63, P65, P66, P68, P71, P73, P74, P75, P76). The identified information primarily aids the attacker in deciding the follow-on actions (P55, P57, P61, P64, P69, P70, P72, P77). The information is usually collected by running various shell commands and native APIs provided in the operating system, making it practically infeasible to detect and defend (P47, P48, P49, P51, P52, P53, P58, P67). 

\begin{table}[]
\centering
\scriptsize
\setlength{\tabcolsep}{2pt}
\caption{Malicious communication using Application Layer Protocol}
\label{tab:categoryCommunication}
\begin{tabular}{p{75mm}cr}
\toprule
\textbf{Pattern}                                                    & \textbf{$\tau$} & \textbf{C} \\ \midrule
(P78) T1041: Exfiltration Over C2 Channel, T1071: Application Layer Protocol & S & 61 \\
(P79) T1071: Application Layer Protocol, T1105: Ingress Tool Transfer     & S          & 6          \\
(P80) T1005: Data from Local System, T1041: Exfiltration Over C2 Channel  & B          & 6          \\
(P81) T1560: Archive Collected Data, T1041: Exfiltration Over C2 Channel  & B          & 5          \\
(P82) T1219: Remote Access Software, T1567: Exfiltration Over Web Service & B          & 3          \\
(P83) T1005: Data from Local System, T1567: Exfiltration Over Web Service & B          & 3          \\
(P84) T1021: Remote Services, T1567: Exfiltration Over Web Service        & B          & 2          \\ \bottomrule
\multicolumn{3}{p{85mm}}{If the command received from the DNS query consists of a string: “uploaddd”, it uploads the local file on the disk using UploadFileAsync() function to an External URL after parsing the TXT record response value into two variables: uriString (external URL) and filename (Local File). This functionality can be leveraged to exfiltrate data.~\cite{zscalaerLyceiumR1025}}
\end{tabular}
\end{table}

\textbf{PC-4: Malicious communication using application layer protocol:}
We identify seven patterns in this category, with 86 occurrences. We report the patterns in Table~\ref{tab:categoryCommunication}. The table shows adversaries use application layer protocols (e.g., HTTP/S, DNS, web services) for both sending (P79) and receiving information and tools (P78, P80-P84). As a result, the traditional network defense and firewall protection rules often do not perform well in detecting this malicious traffic. 

\begin{table}[]
\centering
\scriptsize
\setlength{\tabcolsep}{2pt}
\caption{Lateral movement using OS and Credentials}
\label{tab:categoryLateral}
\begin{tabular}{p{75mm}cr}
\toprule
\textbf{Pattern}                                                        & \textbf{$\tau$} & \textbf{C} \\ \midrule
(P85) T1021: Remote Services, T1570: Lateral Tool Transfer                       & S & 23 \\
(P86) T1003: OS Credential Dumping, T1078: Valid Accounts                     & B          & 8          \\
(P87) T1003: OS Credential Dumping, T1021: Remote Services                    & B          & 6          \\
(P88) T1047: Windows Management Instrumentation, T1570: Lateral Tool Transfer & S          & 6          \\
(P89) T1078: Valid Accounts, T1021: Remote Services                           & B          & 5          \\
(P90) T1021: Remote Services, T1219: Remote Access Software                   & B          & 5          \\
(P91) T1003: OS Credential Dumping, T1570: Lateral Tool Transfer              & B          & 4          \\
(P92) T1047: Windows Management Instrumentation, T1059: Command and Scripting Interpreter & S          & 4           \\
(P93) T1078: Valid Accounts, T1570: Lateral Tool Transfer                     & B          & 3          \\
(P94) T1552: Unsecured Credentials, T1078: Valid Accounts                     & B          & 2          \\
(P95) T1059: Command and Scripting Interpreter, T1570: Lateral Tool Transfer  & S          & 2          \\
(P96) T1569: System Services, T1570: Lateral Tool Transfer                    & B          & 2          \\
(P97) T1110: Brute Force, T1078: Valid Accounts                               & B          & 2          \\
(P98) T1003: OS Credential Dumping, T1550: Use Alternate Authentication Material          & B          & 2           \\
(P99) T1021: Remote Services, T1550: Use Alternate Authentication Material    & B          & 2          \\
(P100) T1003: OS Credential Dumping, T1047: Windows Management Instrumentation & B          & 2          \\ \bottomrule

\multicolumn{3}{p{85mm}}{the operators managed to dump the domain administrator’s NTLM hash. The threat actors then pivoted to the two domain controllers and deployed Cobalt Strike beacons.~\cite{dfirReportZeroToDomainAdminDfir21}}

\end{tabular}
\end{table}

\textbf{PC-5: Lateral movement using OS and Credentials:}
We identify seven patterns in this category, with 86 occurrences in total. We report the patterns in Table~\ref{tab:categoryLateral}. The table shows that adversaries abuse operating system features and credential information to propagate laterally across the victim environment. Attackers use remote services and tools to propagate laterally; for example, they can move to other connected devices using Windows Shared Message Bus or remote desktop protocol (P85, P89, P90, P92). Besides remote services, adversaries abuse Windows Management Instrumentation (P88, P92, P95) and \verb|svchost.exe| to distribute malicious software from domain administrator devices (P96). However, adversaries also need login credentials to move laterally across the victim domain, and adversaries often dump the credential from the LSASS process memory, NTDS, to get the domain controller's credential first (P86, P94, P97). Later, the credentials are used to login to the remotely connected devices (P87, P91, P93, P99, P100), followed by changing the authentication mechanism so that the attackers can maintain the foothold (P98). 

\begin{table}[]
\centering
\scriptsize
\setlength{\tabcolsep}{2pt}
\caption{Importing second stage tools}
\label{tab:categoryImport}
\begin{tabular}{p{75mm}cr}
\toprule
\textbf{Pattern}                                             & \textbf{$\tau$} & \textbf{C} \\ \midrule
(P101) {T1105: Ingress Tool Transfer, T1082: System Information Discovery}  & {B} & {9} \\
(P102) T1105: Ingress Tool Transfer, T1496: Resource Hijacking      & B          & 9          \\
(P103) T1105: Ingress Tool Transfer, T1049: System Network Connections Discovery   & B          & 4          \\
(P104) T1105: Ingress Tool Transfer, T1016: System Network Configuration Discovery & B          & 3          \\
(P105) T1105: Ingress Tool Transfer, T1219: Remote Access Software  & B          & 3          \\
(P106) T1570: Lateral Tool Transfer, T1003: OS Credential Dumping   & B          & 2          \\
(P107) T1105: Ingress Tool Transfer, T1113: Screen Capture          & B          & 2          \\
(P108) T1105: Ingress Tool Transfer, T1547: Boot or Logon Autostart Execution      & B          & 2          \\
(P109) T1105: Ingress Tool Transfer, T1069: Permission Groups Discovery            & B          & 2          \\
(P110) T1105: Ingress Tool Transfer, T1018: Remote System Discovery & B          & 2          \\
(P111) T1105: Ingress Tool Transfer, T1005: Data from Local System  & B          & 2          \\ \bottomrule

\multicolumn{3}{p{85mm}}{Example for P102: The script then downloads and executes XMRig.~\cite{dfirReportCryptoMinerExploitDfir40}}

\end{tabular}
\end{table}

\textbf{PC-6: Importing second stage tools:}
We identify 11 patterns in this category, with 40 occurrences. We report the patterns in Table~\ref{tab:categoryImport}. The table shows that adversaries import second-stage tools for specific malicious operating depending on attack stages: (a) collect victim system-specific information (P101, P103, P104, P109, P110), (b) sensitive data (P106, P107, P111), (c) lateral movement (P105), (d) change system configuration (P108) and (e) mining crypto-currencies (P102). Table~\ref{tab:categoryBaiting} shows that importing second-stage tools usually occurs after the initial compromise, primarily tricking the user into executing malicious code (P2).  

\begin{table}[]
\centering
\scriptsize
\setlength{\tabcolsep}{2pt}
\caption{Exploitation}
\label{tab:categoryExploit}
\begin{tabular}{p{75mm}cr}
\toprule
\textbf{Pattern}                                                                     & \textbf{$\tau$} & \textbf{C} \\ \midrule
(P112) T1190: Exploit Public-Facing Application, T1105: Ingress Tool Transfer      & B & 4 \\
(P113) T1190: Exploit Public-Facing Application, T1505: Server Software Component           & B          & 3          \\
(P114) T1068: Exploitation for Privilege Escalation, T1021: Remote Services                 & B          & 2          \\
(P115) T1210: Exploitation of Remote Services, T1550: Use Alternate Authentication Material & B          & 2          \\ \bottomrule

\multicolumn{3}{p{85mm}}{The threat actor leveraged the WebLogic vulnerability to spawn a command shell from the server running in the Java process, which then in turn, was used to run PowerShell and collect the final payload that the threat actor wished to run on the system.~\cite{dfirReportWeblogicRCEDfir30}}

\end{tabular}
\end{table}

\textbf{PC-7: Exploitation:}
We identify four patterns in this category, with 11 occurrences in total. We report the patterns in Table~\ref{tab:categoryExploit}. The table shows that adversaries exploit vulnerabilities in the victim systems to facilitate follow-on actions. They exploit publicly accessible web and database servers to enable downloading and persisting the compromise (P112, P113). On the other hand, during an active attack, adversaries may exploit vulnerabilities in printers and remote service programs to facilitate lateral movement (P114, P115), primarily print spool service and Windows SMB. 

\begin{table}[]
\centering
\scriptsize
\setlength{\tabcolsep}{2pt}
\caption{Indicator for data breach}
\label{tab:categoryBreach}
\begin{tabular}{p{75mm}cr}
\toprule
\textbf{Pattern}                                            & \textbf{$\tau$} & \textbf{C} \\ \midrule
(P116) T1082: System Information Discovery, T1560: Archive Collected Data & B & 2 \\
(P117) T1114: Email Collection, T1217: Browser Bookmark Discovery  & C          & 2          \\
(P118) T1087: Account Discovery, T1560: Archive Collected Data     & B          & 2          \\
(P119) T1217: Browser Bookmark Discovery, T1555: Credentials from Password Stores  & C          & 2          \\
(P120) T1003: OS Credential Dumping, T1560: Archive Collected Data & B          & 2          \\ \bottomrule

\multicolumn{3}{p{85mm}}{Example for P116: Already covered in a previous case, the batch and PowerShell scripts serve as a data collector to enumerate hosts within the target environment. It collects data about active/dead hosts, disks, and installed software; and stores it in a zip file.~\cite{dfirReportTrickbot1PasswordDfir25}}

\end{tabular}
\end{table}

\textbf{PC-8: Precursor to data-breach:}
We identify five patterns in this category, with ten occurrences in total. We report the temporal patterns in this category in Table~\ref{tab:categoryExploit}. The table shows that adversaries collect information and then stage it in a centralized location through archiving before exfiltration (P116, P118, P120). On the other hand, adversaries look for browser-saved passwords, emails, and bookmarks at the same phase of attacks to collect sensitive information from the logged-in websites of the victim users (P117, P119).

\begin{table}[]
\centering
\scriptsize
\setlength{\tabcolsep}{2pt}
\caption{Indicator for ransomware}
\label{tab:categoryransomware}
\begin{tabular}{p{75mm}cr}
\toprule
\textbf{Pattern}                                            & \textbf{$\tau$} & \textbf{C} \\ \midrule
(P121) T1021: Remote Services, T1486: Data Encrypted for Impact & B & 3 \\
(P122) T1485: Data Destruction, T1486: Data Encrypted for Impact  & B          & 3          \\
(P123) T1486: Data Encrypted for Impact, T1490: Inhibit System Recovery     & B          & 2          \\
(P124) T1567: Exfiltration Over Web Service, T1486: Data Encrypted for Impact  & B          & 2          \\ \bottomrule

\multicolumn{3}{p{85mm}}{Example for P121: For the next hour, the threat actor proceeded to make RDP connections to other servers in the environment. Once the threat actor had a handle on the layout of the domain, they prepared to deploy the ransomware by copying the ransomware (named ttsel.exe ) to each host through the C\$ share folder.~\cite{dfirReportQuantumRansomwareDfir15}}

\end{tabular}
\end{table}

\textbf{PC-9: Precursor to ransomware:}
We identify four patterns in this category, with ten occurrences in total. We report the temporal patterns in this category in Table~\ref{tab:categoryExploit}. The table shows adversaries exfiltrate sensitive data (P124) and disable all system recovery options (P122, P123). On the other hand, P121 indicates that ransomware is distributed from the domain admin device via remote services throughout all the victim devices in the network (P121).


\section{Threats to validity}
In this section, we discuss several limitations of our paper. 
\textbf{Conclusion validity}: In Steps S1C, S2A, and S2B, we built ground truth datasets, which are subject to the first author's judgment. We account for this by including the second, third, and fourth authors in constructing the dataset. Nonetheless, we acknowledge that the dataset may depend on the authors' domain expertise and judgment. \textbf{Content validity:} We acknowledge that the identified temporal patterns depend on the depth of the technical descriptions provided in the report. In Step S1B, we used inclusion and exclusion criteria to select only the CTI reports with acceptable verbosity. However, the filtering led to fewer reports - however, in the literature, several studies~\cite{li2022attackg, husari2017ttpdrill} in extracting CTI and temporal relations have used similar sample sizes. \textbf{Construct validity:} Our work does not investigate all temporal relations specified in TimeML, and techniques cataloged in ATT\&CK. However, we provide a detailed methodology to replicate and engineer the features for temporal relations as well as existing techniques to be introduced in the future versions of ATT\&CK or any other taxonomy. \textbf{Internal validity:} ChronoCTI shows lower recall, indicating that ChronoCTI missed many true positives in Step S3A. However, the higher precision of ChronoCTI results in fewer false alerts. \textbf{External validity:} Our findings may not be subject generalizable in the case of cyberattacks in the long term future. The findings are observed from past cyberattacks, and attackers tend to introduce variety in future attempts. Moreover, we found the patterns from openly available CTI reports, not proprietary sources.

\section{Discussion}
\label{sec:discussion}
In this section, we provide several strategies to mitigate the attack patterns, along with other implications.
\begin{enumerate*}[leftmargin=4mm]
    
    \item \textbf{Cybersecurity best-practices:} We observe from PC-1 that the most prevalent attack pattern is to trick users into executing malicious code. We advocate for educating computer users about cybersecurity best practices in every organization. Moreover, we advocate operating systems and application developers to provide necessary information about cybersecurity best practices for using their products. We advocate users always use tools that can aid in safe browsing - such as malicious URL checkers, strong password recommenders, and password managers. 

    \item \textbf{Immutable OS:} PC-2, PC-5, and PC-9 suggest that adversaries change system configurations to facilitate malicious operations, such as manipulating account settings, injecting malicious code into kernel-level processes, and disabling system recovery. We suggest organizations use immutable operating systems so that no configuration may change even if the system is compromised. Moreover, in PC-4 and PC-6, adversaries install malware after the initial compromise to collect and exfiltrate sensitive data. We advocate practitioners to configure their systems, allowing vendors and administrators to install only signed and verified applications.  

    \item \textbf{Multi user authentication:} PC-2, PC-3, and PC-5 indicate that adversaries can rapidly propagate across the victim network if they can obtain the credentials of domain admins and controllers. As a result, if one admin account is compromised, the whole system becomes compromised instantaneously. To mitigate this, we suggest organizations use multi-user authentications so that in case of a credential breach, attackers may not be able to fully compromise a large number of victim devices quickly. 
    
    \item \textbf{OS with limited functionalities:} PC-2 and PC-5 suggest that several functionalities of operating systems may broaden a system's attack surface, ultimately benefiting the attacker. Examples of such functionalities are credential caching, remote login, and network share - oftentimes, these features are enabled by default - but remains underused by users. We suggest these features be disabled or configured with limited functionality to reduce the attack surface. 
    
    \item \textbf{Read operations must be protected:} PC-3 indicates adversaries use many system-level tools first to identify a victim system's type and specific attributes. While running this program, as a singleton, is not malicious, in the context of an attack, these programs lead attackers to deploy further custom-built malware. We advocate practitioners to restrict the use of these system tools - for example, a command for identifying the system version should only be allowed with root or administrator permission - which would block adversaries from gaining easy access to identify system-level information easily after an initial compromise. 
        
    \item \textbf{Use of encryption and physical backup:} PC-8 and PC-9 suggest that sensitive information stored in victim machines is searched and exfiltrated with sufficient compromise. We encourage organizations to encrypt sensitive information documents, periodic backups on isolated servers, or physical copies.   
            
    \item \textbf{Future research direction:} ChronoCTI has higher precision and lower recall - which motivates further investigation into improving the recall and incorporating all techniques and temporal relations. Moreover, we suggest the introduction of threat models that can specify how actions follow on-another - which would aid practitioners in representing and sharing the chain of attack actions clearly and concisely. We advocate security application developers to devise a threat-hunting mechanism based on the identified patterns. Moreover, alert fatigue is a commonly known issue in intrusion detection~\cite{milajerdi2019holmes} - the identified patterns, paired with machine learning, can be investigated to reduce the number of false alerts. Finally, we also plan to investigate the human factors associated with cyberattacks, as we have observed that the most prevalent patterns of attacks show the victim users have to \textit{act} first to initiate an attack on behalf of the attackers, unknowingly. 
\end{enumerate*}

\section{Related work}
\label{sec:relatedWork}
In this section, we discuss several studies related to our work.  Researchers have contributed to NLP and ML techniques to identify cyberthreat intelligence from CTI reports. Primarily, three types of information are extracted. First, extracting the indicators (such as malware hash, botnet IP address) from text, which is relatively earliest among the three~\cite{zhu2018chainsmith, liu2022tricti}. Second, extraction of tactics, techniques, and procedures from CTI reports~\cite{ge2023explainable, orbinato2022automatic, you2022tim, wu2021price} have been investigated for the past several years - either as a text classification task or information extraction task. Third, of late, cybersecurity-related knowledge graph generation from CTI reports has gained attention from researchers~\cite{ahmed2024cyberentrel, li2022attackg, huang2022building, ren2022cskg4apt, satvat2021extractor}. Researchers have also performed detailed systematic literature reviews on information extraction from CTI reports~\cite{rahman2023attackers, zhao2023survey, rahman2020literature}. The co-occurrence of techniques observed in past cyberattacks has also been investigated in~\cite{shin2023exploiting, al2020learning, rahman2022investigating}. Researchers have utilized the combinations and co-occurrences of adversarial actions to correlate intrusion alert logs, system logs, and kernel API calls to minimize false positives~\cite{chen2024ctimd, gao2021enabling, berady2021ttp, huang2021open, milajerdi2019holmes, milajerdi2019poirot}. In this work, we contribute further to the domain of actionable CTI extraction by mining temporal relations and deriving temporal patterns from CTI reports through proposing and applying \textbf{ChronoCTI} - which is the first pipeline to mine patterns from past cyberattack incidents.

\section{Conclusion}
\label{sec:conclusion}
Security practitioners need to identify high-level adversary behavior to design proactive defense. CTI reports can aid practitioners with proactive defense as the reports contain verbose descriptions of past cyberattacks. The reports are rich in semantic information related to what actions may follow one another or happen simultaneously - which we refer to as the temporal relation between actions. Automatically extracting such temporal relations among attack actions from text is a challenging task and has never been investigated before. To this end, we propose \textbf{ChronoCTI}, an NLP and ML-based pipeline for automatically mining temporal attack patterns from CTI reports. In terms of performance, \textbf{ChronoCTI} does well in precision but lacks recall. Through applying \textbf{ChronoCTI} on 713 CTI reports, we have identified 124 temporal patterns across nine pattern categories. The most prevalent pattern category is such that the adversaries trick users into initiating malicious operations. The identified patterns imply that organizations should assert maximum priority on training general-purpose users on following cybersecurity best practices. We also advocate for adopting immutable systems and multi-user authentications to minimize risks in case of compromises.       


\ifCLASSOPTIONcompsoc
  \section*{Acknowledgments}
\else
  \section*{Acknowledgment}
\fi

This work is partly supported by the NSA Science of Security award H98230-17-D-0080. Any findings and opinions expressed in this material are those of the authors and do not necessarily reflect the views of the funding agencies.

\bibliographystyle{IEEEtran}
\bibliography{main}

\ifCLASSOPTIONcaptionsoff
  \newpage
\fi



\end{document}